\documentclass[hidelinks]{ragtime} % NO optional arguments are required
\title[Levitating atmospheres around naked singularities]%
      {Levitating atmospheres around\\ naked singularities}

\author[R. S. S. Vieira % Run. head authors: separate names with commas,
        and W. Klu\'zniak    % the last one with `and' without a comma.
]%    Now let's start the paper title authors:
       {Ronaldo S. S. Vieira\at[]{1,a} % Makes referencing superscript `1'
        and W\l odek Klu\'zniak\at[]{2,b}\\ % ref. superscr. `1,a',
        \ins{1}Centro de Ci\^encias Naturais e Humanas, Universidade Federal do ABC, 09210-580\splitins[1] Santo Andr\'e, SP, Brazil\\% This is how to break an
        \ins{2}Copernicus Astronomical Center, ul. Bartycka
        18, PL-00-716, Warszawa, Poland\\% Termination of the second affiliation.
        \ins{a}\Email{ronaldo.vieira@ufabc.edu.br} % This is how to present E-mail.
        \ins{b}\Email{wlodek@camk.edu.pl}}

\begin{document}

% Citation of references in abstract should generally be avoided to
% ensure self-consistency of the abstract.  If you do insist on citation(s)
% within the abstract, you should use the \bibentry command, which forces
% the _complete_bibliographic_entry_ to appear in the abstract.
% With the `nonatbib' optional class argument this feature is not available.
\begin{abstract}
For a wide class of spherically symmetric naked singularities there is a sphere within which gravity is effectively repulsive. In such spacetimes accreting matter cannot reach the singularity and will instead form a levitating atmosphere, which is kept suspended by gravity alone. The density of the atmosphere has a maximum at a definite radius. In its qualitative properties the atmosphere is analogous to the recently discussed atmospheres that are supported by radiation pressure above luminous neutron stars, however for the levitating atmospheres around a naked singularity no radiation need be present.
\end{abstract}

% The key words are to be separated by the N-dash surrounded by spaces,
% the left one non-breakable.  The name concatenation like Kerr--de~Sitter
% should also be typed with N-dash but with no spaces around (compare,
% e.g., Levi-Civita, which is a single person):

% It is good to provide as many as \label's possible, but never start the
% key with a numeral.  This makes problems with pdflatex processing.
\section{Introduction}\label{intro}%%%%%%%%%%%%%%%%%%%%%%%%%%%%%%%%%%%%%%%%

Stellar atmospheres are widely studied in the Newtonian gravity context, a general qualitative property being that their density and pressure profiles always decrease with radius. Recently, it was discovered that near-Eddington luminous neutron stars may have atmospheres detached from their surfaces if general relativistic effects are taken into account, either in the optically thin or in the optically thick case \citep{wielgusEtal2015MNRAS, wielgusEtal2016MNRAS}. These equilibrium configurations were called `levitating atmospheres' \citep{wielgusEtal2015MNRAS}, since they lie at a finite distance from the stellar surface and have definite inner and outer radii. The oscillation modes of these atmospheres were studied in \cite{abarcaKluzniak2016MNRAS, bollimpalliKluzniak2017MNRAS, bollimpalliEtalMNRAS2019}. 

Here we investigate whether these structures can appear in the absence of radiation. We find that indeed they can, if we consider spherically symmetric spacetimes generated by a central naked singularity. 
These naked-singularity spacetimes appear not only in general relativity, but also in solutions to modified theories of gravity. They generally present a zero-gravity radius, a stable equilibrium point for radial particle motion. We present below equilibrium atmospheric solutions which `levitate' around these singularities, being supported solely by gravity -- the central repulsive-gravity region.

\section{Repulsive gravity in naked-singularity spacetimes} 	\label{sec:repulsivegravity}%%%%%%%%%%%%%%%%%

Let us consider a spherically symmetric spacetime metric of the form\footnote{We work in geometrized units $c = G = 1$.}
\begin{equation}\label{eq:metric}
ds^2=-e^{2\Phi}\,dt^2 + e^{2\Lambda}\,dr^2 + r^2\,(d\theta^2 + \sin^2\theta\,d\varphi^2)\,. 
\end{equation}
The four-acceleration of a static observer is given by $a_\mu = \partial_\mu\Phi$ \citep{semerakZellerinZacek1999MNRAS}. For metric~(\ref{eq:metric}) we have only a radial component, $a_r = \Phi'(r)$. 
Therefore, if there is a radius $r_o$ such that $\Phi'(r_o) = 0$, particles at rest in the sphere labelled by $r_o$ will remain at rest. We will have then an equilibrium radius for test-particle motion. If $\Phi''(r_o) > 0$ this equilibrium is stable; moreover, for $r<r_o$ we will have a region where gravity behaves as a `repulsive force'. We call the equilibrium sphere of radius $r_o$ the `zero-gravity sphere'. Many spherically symmetric naked-singularity spacetimes behave this way, having an equilibrium radius for particle motion at a finite distance from the singularity (e.g., \citealp{puglieseQuevedoRuffini2011PRD,  stuchlikSchee2014CQG, vieiraMarekEtal2014PRD, katkaVieiraEtal2015GRG, boshkakayev2016PRD}). We will see that, whenever this is the case, a spherical shell of matter will inevitably form around this radius, whose thickness becomes larger as its peak density grows.

\section{Levitating atmospheres}\label{atmospheres} %%%%%%%%%%%%%%%%%%%%%%%%%%%%

If $\Phi''(r_o) > 0$ at the zero-gravity sphere, then we expect that accreting matter onto the singularity starts accumulating around that radius, giving rise to a dense structure with a peak at $r_o$. We call it a levitating atmosphere, in analogy with the recently found levitating atmospheres around luminous neutron stars in general relativity \citep{wielgusEtal2015MNRAS}.

Consider a test-fluid atmosphere around the singularity. It is described by a perfect-fluid energy-momentum tensor
\begin{equation}
T^{\mu\nu} = (\varepsilon + p)\, u^\mu u^\nu + p\,g^{\mu\nu},
\end{equation}
where $\varepsilon$ is the fluid's energy density and $p$ its pressure.
Let us assume that the naked-singularity spacetime is a solution of a theory of gravity where the usual conservation laws are valid for matter.
The equation of hydrostatic equilibrium is then obtained from $T^{\mu\nu}_{\ \ \ ;\nu} = 0$ and, neglecting relativistic contributions to internal energy ($\varepsilon = \rho$, the mass density) and assuming $p\ll \rho$, reads \citep{schutz2009book}
\begin{equation}\label{eq:hydrostaicEquilibrium}
\frac{1}{\rho}\frac{dp}{dr} = - \frac{d\Phi}{dr}.
\end{equation}
This equation has the same form of the Newtonian equation of hydrostatic equilibrium, where $\Phi$ is the gravitational potential.

\subsection{Geometrically thin approximation}

By the preceding discussion, the atmospheres should have a peak of density and pressure at the zero-gravity radius $r_o$. 
If the atmospheres are geometrically thin, we can Taylor expand the function $\Phi$ to second order in $(r-r_o)$ as 
\begin{equation}
	\Phi(r) - \Phi(r_o) \approx \frac{1}{2}\Phi''(r_o)\,(r-r_o)^2
\end{equation}
and then solve the hydrostatic equilibrium equation (\ref{eq:hydrostaicEquilibrium}) given the atmosphere's equation  of state. 

\subsubsection{Isothermal solution}

For an isothermal atmosphere with an ideal gas equation of state 
\begin{equation}
	p = [k_B T/(\mu\, m_p)]\,\rho\,,
\end{equation}
 where $k_B$ is Boltzmann's constant, $m_p$ is the proton mass, and $\mu = 1$ is the mean molecular weight for Hydrogen (since the gas is not ionized), we obtain a Gaussian profile for its pressure, peaked at $r=r_o$, 
\begin{equation}
p(r) = p_o \exp\left[ -\frac{\mu\, m_p}{\,2k_B T\,}\,\Phi''(r_o)\,(r-r_o)^2 \right],
\end{equation}
with the corresponding density profile being given by the ideal gas law. 

\subsubsection{Polytropic solutions}

For a polytropic equation of state of the form 
\begin{equation}
	p = K\rho^\gamma
\end{equation}
we have
\begin{equation}
p(r) = p_o\,\left\{ 1 - 
\frac{1-1/\gamma}{\,2K^{1/\gamma}\,(p_o)^{1-1/\gamma}\,}\, \Phi''(r_o)\,(r-r_o)^2 \right\}^{\gamma/(\gamma-1)},
\end{equation}
with $p_o = p(r_o)$, and a corresponding density profile
\begin{equation}
\rho(r) = \rho_o\left\{ 1 - 
\frac{1-1/\gamma}{\,2K\,\rho_o^{(\gamma-1)}\,}\, \Phi''(r_o)\,(r-r_o)^2 \right\}^{1/(\gamma-1)}.
\end{equation}
The temperature profile is given by
\begin{equation}
T = T_o \left\{1 - \left(\frac{1-1/\gamma}{k_B T_o/(\mu\, m_p)}\right)\cdot\frac{1}{2}\,\Phi''(r_o)\,(r-r_o)^2\right\}^\gamma, 
\end{equation}
assuming an ideal gas law. We remark that the isothermal solution falls exponentially with coordinate distance from $r_o$ in this approximation. However, we do not expect the atmosphere to be isothermal near the singularity, so there should be a cutoff radius at a finite value of $r$. On the other hand, all the  polytropic solutions have (positive) inner and outer rims given by the condition $p(r) = 0$.

Therefore, in the geometrically thin approximation all the profiles depend on the spacetime metric only via two numbers: the zero-gravity radius $r_o$ and the (positive) value of $\Phi''(r_o)$. In particular, the atmospheres will have the same shape regardless of the peculiarities of each spacetime; for a given equation of state, the difference will appear in the position of its peak and in the width of the profiles.

\section{Conclusions}\label{conclus}%%%%%%%%%%%%%%%%%%%%%%%%%%%%%%%%%%%%%%%

If accreting matter gradually falls onto the singularity, for instance via an accretion disc, then it will lose energy while falling the potential well of $\Phi$. It is a property of thin accretion discs in the presence of central repulsive gravity that their inner rim is precisely at the zero-gravity radius (e.g., \citealp{stuchlikSchee2014CQG, vieiraMarekEtal2014PRD}). Once matter is `detached' from the accretion disc and starts falling onto the singularity, it will have to climb the potential well generated by the central repulsive gravity region. Since the energy of the flowing matter will not be sufficient to do so, it will be driven back to larger radii and oscillate around $r_o$. Eventually, due to viscous forces in the fluid, it will settle down at $r_o$ and form the levitating atmosphere.

In this way, matter will never reach the singularity. If enough matter is deposited in the atmosphere, so that the geometrically thin approximation is not valid anymore, exact solutions may give us the optical depth of denser atmospheres and therefore define whether these may be optically thick. In that case, for external observers, the singularity may appear not so different from a gas planet. Therefore a natural astrophysical process, namely accretion, may `cloak' the singularity with a dense spherical layer of optically thick gas. The question of whether this `levitating cloak' also occurs in nonspherical configurations, such as rotating singularities, deserves further investigation.

% Acknowledgements are created using the command \ack:
\ack%%%%%%%%%%%%%%%%%%%%%%%%%%%%%%%%%%%%%%%%%%%%%%%%%%%%%%%%%%%%%%%%%%%%%%%

% Contributors involved in `Vyzkumny zamer' can use macro \InstResCode
% instead of specifying the alphanumerical code explicitly:
The authors thank Maciek Wielgus for interesting discussions. WK acknowledges the hospitality at Federal University of ABC, Brazil, where part of this work was developed. This study was financed in part by the Coordena\c{c}\~ao de Aperfei\c{c}oamento de Pessoal de N\'ivel Superior - Brasil (CAPES) - Finance Code 001, via the Brazilian CAPES - PrInt internationalization program, and by the Polish NCN grant No. 2019/33/B/ST9/01564.

% Here we specify the basename of the bibliography database file,
% in this case \jobname=ragsamp:
%%\bibliography{\jobname}
%\bibliographystyle{h-physrev_titulos}
%\bibliography{ragsamp}
%%%%%\bibliography{refs_RAGtime22}

\begin{thebibliography}{12}
	\expandafter\ifx\csname natexlab\endcsname\relax\def\natexlab#1{#1}\fi
	\expandafter\ifx\csname url\endcsname\relax
	\def\url#1{\texttt{#1}}\fi
	\expandafter\ifx\csname urlprefix\endcsname\relax\def\urlprefix{URL }\fi
	\providecommand{\selectlanguage}[1]{\relax}
	\providecommand{\eprint}[2][]{\url{#2}}
	
	\bibitem[{{Abarca} and {Klu{\'z}niak}(2016)}]{abarcaKluzniak2016MNRAS}
	{Abarca}, D. and {Klu{\'z}niak}, W. (2016), {Radial oscillations of a
		radiation-supported levitating shell in Eddington luminosity neutron stars},
	\emph{Mon. Not. R. Astron. Soc.}, \textbf{461}(3), pp. 3233--3238, \eprint{1604.04485},
	\urlprefix\url{https://ui.adsabs.harvard.edu/abs/2016MNRAS.461.3233A}.
	
	\bibitem[{{Bollimpalli} and
		{Klu{\'z}niak}(2017)}]{bollimpalliKluzniak2017MNRAS}
	{Bollimpalli}, D.~A. and {Klu{\'z}niak}, W. (2017), {Radial modes of levitating
		atmospheres around Eddington luminosity neutron stars}, \emph{Mon. Not. R. Astron. Soc.},
	\textbf{472}(3), pp. 3298--3303, \eprint{1703.04224},
	\urlprefix\url{https://ui.adsabs.harvard.edu/abs/2017MNRAS.472.3298B}.
	
	\bibitem[{{Bollimpalli} et~al.(2019){Bollimpalli}, {Wielgus}, {Abarca} and
		{Klu{\'z}niak}}]{bollimpalliEtalMNRAS2019}
	{Bollimpalli}, D.~A., {Wielgus}, M., {Abarca}, D. and {Klu{\'z}niak}, W.
	(2019), {Atmospheric oscillations provide simultaneous measurement of neutron
		star mass and radius}, \emph{Mon. Not. R. Astron. Soc.}, \textbf{487}(4), pp. 5129--5142,
	\eprint{1812.01299},
	\urlprefix\url{https://ui.adsabs.harvard.edu/abs/2019MNRAS.487.5129B}.
	
	\bibitem[{{Boshkayev} et~al.(2016){Boshkayev}, {Gasper{\'{\i}}n},
		{Guti{\'e}rrez-Pi{\~n}eres}, {Quevedo} and {Toktarbay}}]{boshkakayev2016PRD}
	{Boshkayev}, K., {Gasper{\'{\i}}n}, E., {Guti{\'e}rrez-Pi{\~n}eres}, A.~C.,
	{Quevedo}, H. and {Toktarbay}, S. (2016), {Motion of test particles in the
		field of a naked singularity}, \emph{Phys. Rev. D}, \textbf{93}(2), 024024,
	\eprint{1509.03827},
	\urlprefix\url{http://adsabs.harvard.edu/abs/2016PhRvD..93b4024B}.
	
	\bibitem[{{Goluchov{\'a}} et~al.(2015){Goluchov{\'a}}, {Kulczycki}, {Vieira},
		{Stuchl{\'{\i}}k}, {Klu{\'z}niak} and {Abramowicz}}]{katkaVieiraEtal2015GRG}
	{Goluchov{\'a}}, K., {Kulczycki}, K., {Vieira}, R.~S.~S., {Stuchl{\'{\i}}k},
	Z., {Klu{\'z}niak}, W. and {Abramowicz}, M. (2015), {Ho{\v r}ava's quantum
		gravity illustrated by embedding diagrams of the Kehagias-Sfetsos
		spacetimes}, \emph{General Relativity and Gravitation}, \textbf{47}, 132,
	\eprint{1511.01345},
	\urlprefix\url{http://adsabs.harvard.edu/abs/2015GReGr..47..132G}.
	
	\bibitem[{{Pugliese} et~al.(2011){Pugliese}, {Quevedo} and
		{Ruffini}}]{puglieseQuevedoRuffini2011PRD}
	{Pugliese}, D., {Quevedo}, H. and {Ruffini}, R. (2011), {Circular motion of
		neutral test particles in Reissner-Nordstr{\"o}m spacetime}, \emph{Phys. Rev. D},
	\textbf{83}(2), 024021, \eprint{1012.5411},
	\urlprefix\url{http://adsabs.harvard.edu/abs/2011PhRvD..83b4021P}.
	
	\bibitem[{Schutz(2009)}]{schutz2009book}
	Schutz, B. (2009), \emph{A first course in general relativity}, Cambridge
	university press, UK.
	
	\bibitem[{{Semer{\'a}k} et~al.(1999){Semer{\'a}k}, {Zellerin} and {{\v
				Z}{\'a}{\v c}ek}}]{semerakZellerinZacek1999MNRAS}
	{Semer{\'a}k}, O., {Zellerin}, T. and {{\v Z}{\'a}{\v c}ek}, M. (1999), {The
		structure of superposed Weyl fields}, \emph{Mon. Not. R. Astron. Soc.}, \textbf{308}, pp.
	691--704, \urlprefix\url{http://adsabs.harvard.edu/abs/1999MNRAS.308..691S}.
	
	\bibitem[{{Stuchl{\'{\i}}k} and {Schee}(2014)}]{stuchlikSchee2014CQG}
	{Stuchl{\'{\i}}k}, Z. and {Schee}, J. (2014), {Optical effects related to
		Keplerian discs orbiting Kehagias{\&}Sfetsos naked singularities},
	\emph{Classical and Quantum Gravity}, \textbf{31}(19), 195013,
	\eprint{1402.2891},
	\urlprefix\url{http://adsabs.harvard.edu/abs/2014CQGra..31s5013S}.
	
	\bibitem[{{Vieira} et~al.(2014){Vieira}, {Schee}, {Klu{\'z}niak},
		{Stuchl{\'{\i}}k} and {Abramowicz}}]{vieiraMarekEtal2014PRD}
	{Vieira}, R.~S.~S., {Schee}, J., {Klu{\'z}niak}, W., {Stuchl{\'{\i}}k}, Z. and
	{Abramowicz}, M. (2014), {Circular geodesics of naked singularities in the
		Kehagias-Sfetsos metric of Ho{\v r}ava's gravity}, \emph{Phys. Rev. D},
	\textbf{90}(2), 024035, \eprint{1311.5820},
	\urlprefix\url{http://adsabs.harvard.edu/abs/2014PhRvD..90b4035V}.
	
	\bibitem[{{Wielgus} et~al.(2015){Wielgus}, {Klu{\'z}niak}, {S\c{a}dowski},
		{Narayan} and {Abramowicz}}]{wielgusEtal2015MNRAS}
	{Wielgus}, M., {Klu{\'z}niak}, W., {S\c{a}dowski}, A., {Narayan}, R. and
	{Abramowicz}, M. (2015), {Stable, levitating, optically thin atmospheres of
		Eddington-luminosity neutron stars}, \emph{Mon. Not. R. Astron. Soc.}, \textbf{454}(4), pp.
	3766--3770, \eprint{1505.06099},
	\urlprefix\url{https://ui.adsabs.harvard.edu/abs/2015MNRAS.454.3766W}.
	
	\bibitem[{{Wielgus} et~al.(2016){Wielgus}, {S{\c{a}}dowski}, {Klu{\'z}niak},
		{Abramowicz} and {Narayan}}]{wielgusEtal2016MNRAS}
	{Wielgus}, M., {S{\c{a}}dowski}, A., {Klu{\'z}niak}, W., {Abramowicz}, M. and
	{Narayan}, R. (2016), {Levitating atmospheres of Eddington-luminosity neutron
		stars}, \emph{Mon. Not. R. Astron. Soc.}, \textbf{458}(4), pp. 3420--3428, \eprint{1512.00094},
	\urlprefix\url{https://ui.adsabs.harvard.edu/abs/2016MNRAS.458.3420W}.
	
\end{thebibliography}

%###################################
%###################################

%###################################
%###################################

\end{document}